\documentclass[copyright,creativecommons]{eptcs}
\providecommand{\event}{}
\providecommand{\volume}{}
\providecommand{\anno}{2009}
\providecommand{\firstpage}{1}

\usepackage[dvips]{color}
\usepackage{graphicx}
\usepackage{amssymb}
\usepackage{multicol}
\usepackage{mathrsfs}
\usepackage{stmaryrd}         % semantic brackets
\usepackage{breakurl}
\usepackage{array}
\usepackage{prettyref}

\def\causes #1{{\lfloor} #1 {\rfloor}}
\def\LES #1{{\mathsf{LES}} {\llbracket}  #1 {\rrbracket}}

\definecolor{rdxbackcolor}{gray}{0.91}

\def\multiset  #1{  \{\!\!| \, #1 \,   |\!\!\}}

\def\muscup {\dot\cup} %Multiset union 
\def\musminus{\dot{-}} %Multiset minus

\def\mussubseteq{\,\dot\subseteq\,} %Multiset subseteq
\newcommand{\reduce}{\longrightarrow}
\newcommand{\transition}[1]{\stackrel{#1}{\reduce}}

\newtheorem{theorem}{Theorem}
\newtheorem{definition}[theorem]{Definition}
\newtheorem{example}[theorem]{Example}

\newtheorem{proposition}[theorem]{Proposition}

\global\long\def\reduce{\longrightarrow}
\global\long\def\transition#1{\stackrel{#1}{\reduce}}

\global\long\def\rate#1{\rho(#1)}

\title{Flux Analysis in Process Models via Causality}
\author{Ozan Kahramano\u gullar{\i} 
\institute{%%%%%%%%%
The Microsoft Research -- University of Trento\\ 
Centre for Computational and Systems Biology}
}
\def\titlerunning{
Flux Analysis in Process Models via Causality
}
\def\authorrunning{
{Kahramano\u gullar{\i}}
}
\begin{document}
\maketitle

\begin{abstract}
We present an approach for flux analysis in process algebra models 
of biological systems. We perceive flux as the 
flow of resources in stochastic simulations. We resort 
to an established correspondence between event structures,
a broadly recognised model of concurrency, and state 
transitions of process models, seen as Petri nets.
We show that we can this way extract the 
causal resource dependencies in simulations between individual state 
transitions as partial orders of events. We propose transformations on 
the partial orders that provide means for further analysis, 
and introduce a software tool, which implements these ideas. 
By means of an 
example of a published model of the Rho GTP-binding proteins, we argue 
that this approach can provide the substitute for flux analysis techniques 
on ordinary differential equation models within the stochastic 
setting of process algebras.
\end{abstract}

%%%%%%%%%%%%%%%%

\section{Introduction}

Computer science methodologies are now gaining increasing 
attention in %the field of 
systems biology, where they 
are used together with techniques from applied mathematics 
and physics to provide insights into the workings of 
biological systems. In such a setting, it is often desirable 
to model smaller biological systems as sets of differential 
equations.
Being equipped with well understood analysis techniques, 
differential equations provide the 'deterministic' 
approach (as opposed to the stochastic approach)
for modelling and simulating biological systems. 
Differential equation models, however, are inherently 
difficult to change, extend and upgrade 
when modification to the model structure is required. 
In addition, within the
last few years, a general consensus has emerged that
noise and stochastic effects, which are not directly 
captured by differential equations, are essential 
attributes of biological systems
(see, e.g., \cite{SS08}).

`Tailored to fit' constructs of 
computer science languages provide the means for building 
models of larger systems, especially when it is 
required to refine, modify or compose the models, 
for example when new data is acquired.  
Various languages with stochastic simulation capabilities 
based on, for example, 
process algebras \cite{PRSS01,CZ08b,GDUHP09,PQR09}, term
rewriting (see, e.g, \cite{DFFHK07,FDKHF08}) 
and Petri nets (see, e.g., \cite{HGD08,Tal08}) have been proposed. 
%%%%%%%%%%%%%%%%%%%%%%%%%%%%%%%%%%%%%%%%%%%%%%%%%%%%%%%%%%%%%%%%%%
In particular, process algebras, which were originally introduced 
to study the properties of complex, reactive, concurrent  
systems are well suited for modelling biological systems. By 
using these languages, we can build models to describe the dynamics of the 
biological systems, which are inherently complex and massively 
parallel, and run stochastic simulations. 
The time series data resulting from these stochastic 
simulations are then  used to observe the emergent 
behaviour, together with other techniques that are borrowed 
from theoretical computer science to explore the model for 
hypothesis generation. However, 
these techniques are still underdeveloped in comparison 
to the rich arsenal of differential equation analysis
techniques 
that have their roots in Newton's physics. There is now a 
pressing need for the adjustment of the computer science 
analysis techniques on models of biological systems 
and on stochastic simulations with them so that biologically 
meaningful observations can be made.
 
Along these lines, we present an approach for the analysis
of stochastic simulations with process algebra models.  
The stochastic semantics of process algebras are
conveniently given by continuous time Markov chains (CTMC).
Each simulation is a trajectory of a CTMC, which is 
a sequence of 
computations of the underlying transition system, 
given by the process algebra model. 
The CTMC semantics, thus, 
imposes a total order on the transitions of a 
simulation  trajectory that is emphasised 
by the unique time stamps of the individual transitions.
In this respect, a simulation on a model can be seen as 
reduction of a complex structure, i.e., the model,
into a simpler structure, i.e., the simulation 
trajectory. However, during this reduction some 
of the information on the model is lost, and others 
are made implicit. 
%%%%%%%%%%%%%%%%%%%%%%%%%%%%%%%%%%%
The idea here is to recover this 
implicit information: when these transitions are 
inspected from the point of view of their dependencies on one 
another, it is possible to relax the total order of the transitions 
into a partial order. We can then use this partial order as a representation 
of causal dependencies in the simulation and process it to observe 
the flux in the system with respect to the flow of the resources 
during simulation from one transition to another.

Partial orders reflecting dependencies in  computations 
have been studied as non-interleaving models
of concurrency: when there are no resource conflicts
between two partially ordered events, they can take
place in parallel, and their common predecessors
and successors provide a synchronisation mechanism.
Event structures reflect this view \cite{NPW81,WN95}: 
in an event structure dependencies of events of a concurrent 
system are given with a partial order. In \cite{Kah09}, 
for sequences of computations in Petri nets, 
we have given an algorithm for extracting 
partial orders that exhibit event structure 
semantics\footnote{The algorithm in \cite{Kah09} 
is presented on multiplicative exponential 
linear logic encodings of Petri nets.}. 
In the following, we carry these ideas to CTMCs of process algebra 
simulations by tracing the 'consumption' and 'production' 
effect of each event. 
This way, we extract the information on the flow of resources 
in simulations as partial orders. We then apply
transformations  to these partial orders for flux 
analysis. We illustrate these ideas first on a simple 
example, and then on a published model of Rho GTP-binding 
proteins \cite{GP06,CCGKP08}.

We restrict the presentation in this paper 
to the models that are given as sets of chemical reactions, 
and we use the stochastic $\pi$ calculus and its implementation 
language $\mathsf{SPiM}$ \cite{PL07}. However, we believe that the 
ideas presented in this paper are applicable for a more general 
class of models and to a broader spectrum of modelling languages.

%%%%%%%%%%%%%%%%%%%%%%%%%%%%%%%%%%%%%%%%%%%%%%%%%%%%%%%%%%%%%%%%%%%%%%%%%%%
%%%%%%%%%%%%%%%%%%%%%%%%%%%%%%%%%%%%%%%%%%%%%%%%%%%%%%%%%%%%%%%%%%%%%%%%%%%
%%%%%%%%%%%%%%%%%%%%%%%%%%%%%%%%%%%%%%%%%%%%%%%%%%%%%%%%%%%%%%%%%%%%%%%%%%%
%%%%%%%%%%%%%%%%%%%%%%%%%%%%%%%%%%%%%%%%%%%%%%%%%%%%%%%%%%%%%%%%%%%%%%%%%%%
%%%%%%%%%%%%%%%%%%%%%%%%%%%%%%%%%%%%%%%%%%%%%%%%%%%%%%%%%%%%%%%%%%%%%%%%%%%
%%%%%%%%%%%%%%%%%%%%%%%%%%%%%%%%%%%%%%%%%%%%%%%%%%%%%%%%%%%%%%%%%%%%%%%%%%%
%%%%%%%%%%%%%%%%%%%%%%%%%%%%%%%%%%%%%%%%%%%%%%%%%%%%%%%%%%%%%%%%%%%%%%%%%%%

\section{Processes of Biology}
In this section, we describe how stochastic $\pi$ calculus can be used to 
model biological systems, e.g., as systems of chemical reactions \cite{CCGKP08} 
or as systems of entities with more complex structures, e.g., polymers \cite{CCGKP08b}.
% or membranes \cite{CCCCCC}.  
In stochastic $\pi$ calculus, the basic building blocks are 
processes which are defined as follows.

\begin{definition} \label{spi:syntax}
\cite{CCGKP08} Syntax of the stochastic $\pi$ calculus: 
\emph{processes} range over $P,Q,\ldots$
Below fn$(P)$ denotes the set of names that are free in $P$.

~

\begin{centering}
\begin{tabular}{rc>{\raggedright}p{2cm}rcl}
\texttt{P},\texttt{Q}::= & \texttt{M} & Choice & \texttt{M}::= & \texttt{()} & Null\tabularnewline
| & \texttt{X(n)} & Instance & | & \texttt{$\pi$; P} & Action\tabularnewline
| & \texttt{P | Q} & Parallel & | & \texttt{do $\pi$1;P1 or...or $\pi$N;PN} & Actions\tabularnewline
| & \texttt{new x P} & Restriction &  &  & \tabularnewline
 &  &  & $\pi$::= & \texttt{?x(m){*}r} & Input\tabularnewline
\texttt{E}::= & \texttt{\{\}} & Empty & | & \texttt{!x(n){*}r} & Output\tabularnewline
| & \texttt{E,X(m) = P} & Definition, 

fn(P) $\subseteq$ m & | & \texttt{delay@r} & Delay\tabularnewline
\end{tabular}
\par
\end{centering}

~

Expressions above are considered equivalent up to the least congruence relation 
given by the equivalence relation $\equiv$ defined as follows.

~

\begin{centering}
\texttt{\qquad}\begin{tabular}{crcl}
 & \texttt{P | ()} & \texttt{$\equiv$} & \texttt{P}\tabularnewline
 & \texttt{P | Q} & \texttt{$\equiv$} & \texttt{Q | P}\tabularnewline
 & \texttt{P | (Q | R)} & \texttt{$\equiv$} & \texttt{(P | Q) | R}\tabularnewline
\texttt{X(m) = P} & \texttt{X(n)} & \texttt{$\equiv$} & \texttt{P\{m:=n\}}\tabularnewline
 & \texttt{new x ()} & \texttt{$\equiv$} & \texttt{()}\tabularnewline
 & \texttt{new x new y P} & \texttt{$\equiv$} & \texttt{new y new x P}\tabularnewline
\texttt{x$\notin$fn(P)} & \texttt{new x (P | Q)} & \texttt{$\equiv$} & \texttt{P | new x Q}\tabularnewline
\end{tabular}
\end{centering}
%%%%%%%%%%%%%%%%%%%%%%%%%%%%%%%%%%%%%%%%%%%%%%%%%%%%%%%%%%%%%%%%%%
%%%%%%%%%%%%%%%%%%%%%%%%%%%%%%%%%%%%%%%%%%%%%%%%%%%%%%%%%%%%%%%%%%
\end{definition}

~

While implementing models in the stochastic $\pi$ calculus, 
we follow the abstraction of the biochemical species as entities
that can change state. The state changes can be spontaneous or result from 
the interactions of the species with other species.  
We denote different states of the species with processes. 
Because the state of a species is a process with different
state transition capabilities, a process \texttt{P} can 
choose stochastically between zero or more alternative behaviours. 
In the language of \textsf{SPiM}, we write the choice of $N$ processes 
as  \texttt{do P1 or ... or PN}. A choice of only one process 
is written as \texttt{P1}, while the empty choice is written as 
\texttt{()}. A parallel composition of $N$ processes is written 
as \texttt{P1 | ... | PN}. A process can also be given a name \texttt{X} with 
parameter \texttt{m}, written \texttt{let X(m) = P}.
By using  these constructs, 
we can compose processes to construct models and incrementally extend 
them to build larger models when desired.

The reduction rules of the calculus are as follows. 

\begin{definition}
\label{spi:reduction}
\cite{CCGKP08} Reduction in the stochastic $\pi$ calculus. 

~

%%%%%%%%%%%%%%%%%%%%%%%%%%%%%%%%%%%%%%%%%%%%%%%%%%%%%%%%%%%%%%%%%%
%%%%%%%%%%%%%%%%%%%%%%%%%%%%%%%%%%%%%%%%%%%%%%%%%%%%%%%%%%%%%%%%%%
\begin{centering}
\begin{tabular}{cc>{\raggedleft}p{5cm}cl}
(1) &  & \texttt{do delay@r; P or ... } & \texttt{$\transition r$} & \texttt{P}\tabularnewline
(2) &  & \texttt{(do !x(n){*}r1; P1 or...)}

\texttt{| (do ?x(m){*}r2; P2 or...)} & \texttt{$\transition{\rate x\cdot r1\cdot r2}$} & \texttt{P1 | P2\{m:=n\}}\tabularnewline
(3) & \texttt{P$\transition r$P'} & \texttt{new x P} & \texttt{$\transition r$} & \texttt{new x P'}\tabularnewline
(4) & \texttt{P$\transition r$P'} & \texttt{P | Q} & \texttt{$\transition r$} & \texttt{P' | Q}\tabularnewline
(5) & \texttt{Q$\equiv$P$\transition r$P'$\equiv$Q'} & \texttt{Q} & \texttt{$\transition r$} & \texttt{Q'}\tabularnewline
\end{tabular}
\par\end{centering}
%%%%%%%%%%%%%%%%%%%%%%%%%%%%%%%%%%%%%%%%%%%%%%%%%%%%%%%%%%%%%%%%%%
%%%%%%%%%%%%%%%%%%%%%%%%%%%%%%%%%%%%%%%%%%%%%%%%%%%%%%%%%%%%%%%%%%
\end{definition}

~

A species can spontaneously change state 
after a delay or as a result of a stochastic interaction with another species. 
The stochastic behaviour of the processes is delivered by the rate 
of the delay or interaction actions that the processes can perform:
a process can perform a delay at rate $r$ and then do \texttt{P},
written \texttt{delay@r;P}. The rate $r$ is a real number value 
denoting the rate of an exponential distribution. A process can 
also send a value $n$ on channel $x$ with weight $w_1$ and 
then  do $P_1$, written \texttt{!x(n)*w1;P1}, or it can 
receive a value $m$ on channel $x$ with weight $w_2$ and then 
do $P_2$, written \texttt{?x(m)*w2;P2}. With respect to the reduction 
semantics of \textsf{SPiM}, these complementary send and 
receive actions can synchronise on the common channel $x$ and evolve 
to \texttt{P1 | P2\{m := n\}}, where $m$ is replaced 
by $n$ in process $P_2$.
This allows messages to be exchanged from one process to another.
Each channel name $x$ is associated with an underlying rate given 
by $\rho(x)$. The resulting rate of the interaction is given 
by $\rho(x)$ times the weights $w_1$ and $w_2$.
If no weight is given then a default weight of $1$ is
used. The operator \texttt{new x@r:t P} creates a fresh channel $x$ 
of rate $r$ to be used in the process $P$, where $t$
is the type of the channel. 
When a process is prefixed with the declaration of a fresh channel, 
that channel remains private to the process and does not conflict 
with any other channel. By using these constructs, we can describe the 
dynamics of a biological system as a model 
in the language of \textsf{SPiM}.

\begin{figure}[t]
\begin{tabular}{ccc}
%%%%%%%%%%%%%%%%%%%%%%%%%%%%%%%%%%%%
%%%%%%%%%%%%%%%%%%%%%%%%%%%%%%%%%%%%
\begin{tabular}{c}
\includegraphics[width=30mm]{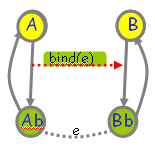} 
\end{tabular}
& 
%%%%%%%%%%%%%%%%%%%%%%%%%%%%%%%%%%%%
%%%%%%%%%%%%%%%%%%%%%%%%%%%%%%%%%%%%
\begin{tabular}{l}
{\footnotesize \texttt{new bind@r1:chan(chan)}}\\
{\footnotesize \texttt{let A() = ( new e@r2:chan}}\\
{\footnotesize \texttt{!bind(e); Ab(e) )}}\\
{\footnotesize \texttt{and Ab(e:chan) =  !e; A()}}\\
{\footnotesize \texttt{let B() = ( ?bind(e); Bb(e) )}}\\  
{\footnotesize \texttt{and Bb(e:chan) =  ?e; B()}} 
\end{tabular}
%%%%%%%%%%%%%%%%%%%%%%%%%%%%%%%%%%%%
%%%%%%%%%%%%%%%%%%%%%%%%%%%%%%%%%%%%
\quad & \quad 
%%%%%%%%%%%%%%%%%%%%%%%%%%%%%%%%%%%%
%%%%%%%%%%%%%%%%%%%%%%%%%%%%%%%%%%%%
\begin{tabular}{l}
{\footnotesize \texttt{new bind@r1:chan()}}\\
{\footnotesize \texttt{let A() = !bind; AB()}}\\  
{\footnotesize \texttt{and AB() =  delay@r2; ( A() | B() )}}\\ 
{\footnotesize \texttt{and B() = ( ?bind(); () )}}
\end{tabular}
%%%%%%%%%%%%%%%%%%%%%%%%%%%%%%%%%%%%
%%%%%%%%%%%%%%%%%%%%%%%%%%%%%%%%%%%%
\end{tabular}
\caption{Two different encodings in \textsf{SPiM} of a model 
of the association of the species $A$ and $B$ to form a complex, 
and their dissociation: 
on the left, we have a model and its graphical representation, 
where the usage of the private name 
\texttt{e} for the bond between the 
species \texttt{Ab} and \texttt{Bb}
permits the monitoring of the individual species in
the complex after the association. 
The model on the right 
abstracts away from this information 
within a simpler representation.
}  
\label{figure:complexation}
\end{figure}

\begin{example}
Consider the biochemical species 
\texttt{A} and \texttt{B} which can associate with rate $r_1$ 
to form a complex 
and dissociate with rate $r_2$. 
In Figure \ref{figure:complexation}, 
we depict two different encodings of this model in \textsf{SPiM}.
\end{example}
For an in depth exposure to the syntax and semantics of the 
stochastic $\pi$ calculus that is implemented in \textsf{SPiM}
and examples of models with varying structures, 
we refer to  \cite{PL07,CCGKP08,Car08c,BCP08,CCGKP08b,KC09}.

%%%%%%%%%%%%%%%%%%%%%%%%%%%%%%%%%%%%%%%%%%%%%%%%%%%%%%%%%%%%%%%%
%%%%%%%%%%%%%%%%%%%%%%%%%%%%%%%%%%%%%%%%%%%%%%%%%%%%%%%%%%%%%%%%
%%%%%%%%%%%%%%%%%%%%%%%%%%%%%%%%%%%%%%%%%%%%%%%%%%%%%%%%%%%%%%%%
%%%%%%%%%%%%%%%%%%%%%%%%%%%%%%%%%%%%%%%%%%%%%%%%%%%%%%%%%%%%%%%%

\section{Causality as Partial Order}
\label{section:causality} 

The stochastic semantics of the $\pi$ calculus models in 
\textsf{SPiM} are given with continuous time Markov chains 
(CTMC). These CTMCs are infinite sets that collect the  
possible simulation trajectories of the modelled systems. 
In each possible trajectory of a process model, 
the duration of each state 
change is delivered by an exponential 
distribution from a continuous interval. Then, a simulation 
with a model is essentially a sequence of state transitions 
of the underlying model, where each transition is stamped with 
a time constant that imposes a total order on these transitions. 
In order to see this on an example, let us first define formally 
the class of models that we discuss in this paper.

\begin{definition}[model]
Biochemical species are denoted with $A, B, P, Q, \ldots$. 
A model $\mathcal{M}$ is a pair $(\mathcal{I},\mathcal{R})$ where
$\mathcal{I}$ is a 
multiset\footnote{Multisets are denoted by      
  the curly brackets ``$\multiset{ \;\;}$''.
 $\muscup\,$,  $\musminus$ and  $\mussubseteq$    denote the 
 multiset operations corresponding to the usual set operations
 $\cup\,$, $-\,$ and  $\subseteq\,$, respectively.} 
%%%%%%%%%%%%%%%%%%%%%%%%%%%%%%%%%%%%%%%%%%%%%%%%%%%%
of species denoting the initial 
state of the model and  $\mathcal{R}$ is a finite 
set of chemical reactions of the form
$$
\mathsf{r}: A  \rightarrow_{\rho} P_1 + \ldots + P_k
\qquad \textrm{or} \qquad
\mathsf{r}: A + B \rightarrow_{\rho} P_1 + \ldots + P_k     
$$
such that $k \in \mathbb{N}$, 
$\mathsf{r}$ is the name of the rule, 
and $\rho \in \mathbb{R}^+$ is the rate of the reaction. 
\end{definition}

\begin{figure}[b]
\begin{tabular}{ccccc}
\begin{tabular}{l}
{\texttt{0.  --> A()}}\\
{\texttt{0.460566394485 A()  --> P() P()}}\\ 
{\texttt{0.734200976191 P()  --> C()}}\\ 
{\texttt{1.13404290455 P()  --> B()}}\\ 
{\texttt{1.86405154022 B() C()  --> D()}}
\end{tabular}
&
\begin{tabular}{l}
$\leadsto$
\end{tabular}
&
\begin{tabular}{l}
\includegraphics[width=30mm]{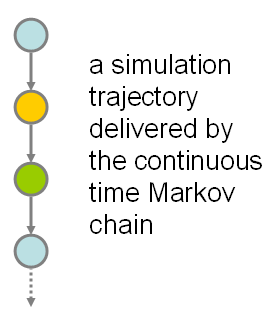}
\end{tabular}
&
\begin{tabular}{l}
$\!\!\!\!\!\!\!\!\!\!\!\!\leadsto\!\!\!\!\!\!\!\!\!\!\!\!\!\!\!\!\!\!\!\!\!\!\!\!\!\!$
\end{tabular}
&
\begin{tabular}{l}
\includegraphics[width=30mm]{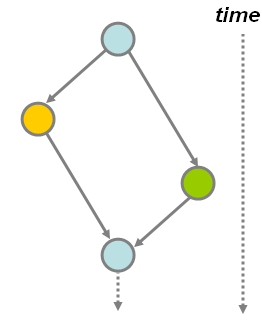}
\end{tabular}
\end{tabular}
\caption{Graphical representation of the extraction of a partial order 
that reflects the production/consumption relationship of the transitions 
in a simulation with the model in Example \ref{example:deriving:po}.}
\label{figure:diamond}
\end{figure}

\begin{example}
\label{example:diamond}
Consider the model with the reactions below and the initial state 
$\mathcal{I} = \multiset{A}$, which can be implemented in   
\textsf{SPiM} as depicted in Figure \ref{figure:spim:model}. 
On the left-hand-side in Figure 
\ref{figure:diamond}, 
we have a simulation output with this model. 
$$
             \mathsf{r}_1: A      \rightarrow_{1.0}  P + P 
\qquad \quad \mathsf{r}_2: P      \rightarrow_{1.0}  B 
\qquad \quad \mathsf{r}_3: P      \rightarrow_{1.0}  C
\qquad \quad \mathsf{r}_4: B + C  \rightarrow_{1.0}  D  
$$
\end{example}

A simulation on such a process model
can be seen as the interleavings of the computations of a 
place/transition Petri net, which can be infinite, e.g., in the case of 
polymerisation models \cite{CCGKP08b,CZ08}: 
in Petri nets, each transition has incoming arrows from places, containing 
tokens that are consumed as a result of the transition, and it has outgoing 
arrows to places, containing the tokens that are produced as a result of the 
transition. Then, each `\texttt{delay}' action is a transition with a 
single incoming place, and each action resulting from the 
interaction of input (`\texttt{?}') and output (`\texttt{!}') 
actions is a transition with two incoming places 
for the input and output actions. 
For the purpose of this paper, we restrict the models to those, which 
can be expressed as finite sets of chemical reactions.
This restriction allows us to denote each model interchangeably as a finite set of chemical 
reactions as well as the corresponding Petri net or the (equivalence class of the)
$\mathsf{SPiM}$ implementation (since there does not exist a unique implementation 
for each set of reactions in $\textsf{SPiM}$).

\begin{figure}[t]
$$
\includegraphics[width=90mm]{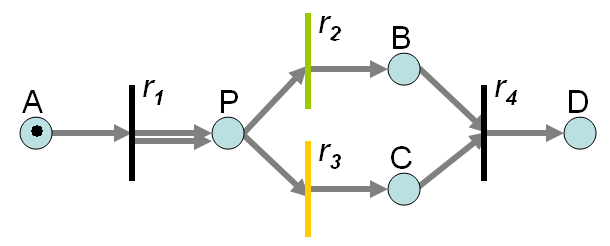}
\qquad 
\includegraphics[width=55mm]{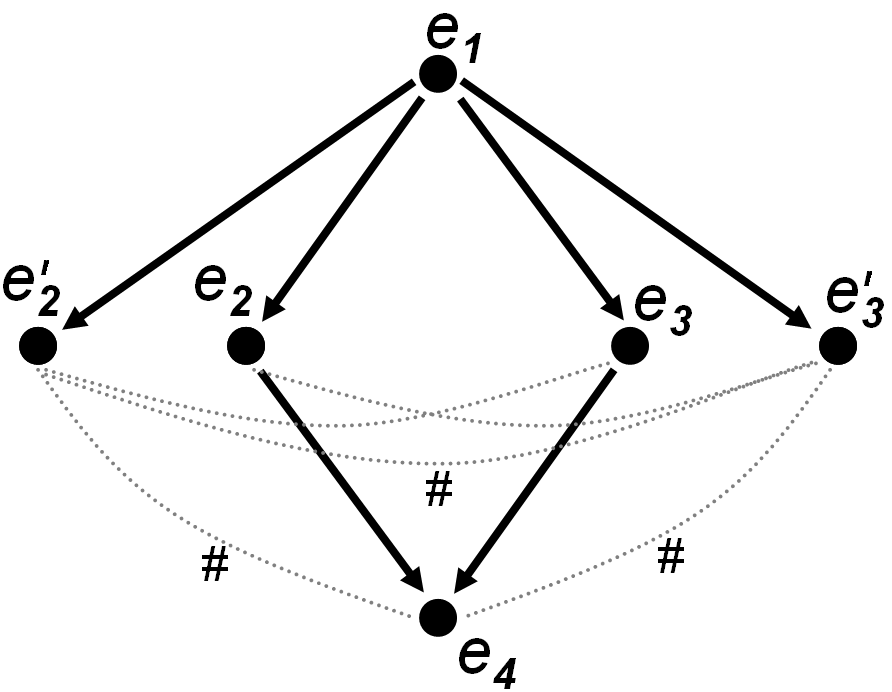}
$$
\caption{The Petri net of the model given in Example \ref{example:diamond} 
and the LES of this model, where the subscripts of the events correspond 
to the subscripts 
of the reactions(transitions) that they refer to. The dashed lines denote the 
conflict relation $\#$.}
\label{figure:Petri}
\end{figure}

\begin{example}
Consider the Petri net depicted in Figure \ref{figure:Petri}, 
which is the net underlying the model 
given in Example \ref{example:diamond}
and implemented in Figure \ref{figure:spim:model}. 
\end{example}

The consideration of a simulation trajectory of a process model 
from the point of view of Petri nets
remains in agreement also with the CTMC semantics that 
imposes a total order on the transitions (see, e.g., \cite{MBCDF95}). 
However, when these transitions are inspected from the point 
of view of their dependencies on one another with respect 
to their production/consumption relationships, it is possible 
to relax this total order into a partial order by abstracting 
away from the time stamps of the state transitions.

\begin{figure}[t]
\begin{multicols}{3}
{\footnotesize
\begin{verbatim}
directive sample 1000.0
directive plot A(); P(); 
               B(); C(); D() 

val r1 = 1.0    val r2 = 1.0      
val r3 = 1.0    val r4 = 1.0      

  new ch@r4:chan()

  let A() = delay@r1; (P() | P())

  and P() = do delay@r2; B()
            or delay@r3; C()

      and B() = ?ch; D()
      and C() = !ch; ()

      and D() = ()

      run 1 of A()  
\end{verbatim}
}
\end{multicols}
\caption{A \textsf{SPiM} implementation of the model given in Example \ref{example:diamond}.}
\label{figure:spim:model}
\end{figure}

\begin{example}
\label{example:deriving:po}
Consider the model given in Example \ref{example:diamond}. 
In a simulation on this model, the reactions 
$\mathsf{r}_2$ and $\mathsf{r}_3$ are totally ordered due to 
the time stamps on them. However, we can relax this 
total order into a partial order with respect to the 
production/consumption relationship between the 
transitions as depicted in Figure \ref{figure:diamond}.  
\end{example}

In fact, the underlying Petri net of a process model hints this partial 
order structure on the transitions, 
as it can be seen on the example net depicted in Figure \ref{figure:Petri}. 
However, the possible conflicts in simulations between various transitions  
with respect to the availability of resources remain implicit in the structure 
of a net. For example, consider a simulation state with 
the Petri net in Figure \ref{figure:Petri}, where there is a
token in place $\texttt{P}$,  and thus there is a conflict between the 
transitions $\mathsf{r}_2$ and $\mathsf{r}_3$. Although
this conflict can be observed by inspecting multiple outgoing arrows 
from places, it is not always explicit 
in the structure of more complex Petri nets.

Partial orders reflecting dependencies in computations 
have been studied as non-interleaving models
of concurrency: in this setting, the synchronisation of 
concurrent transitions is realized 
by their common successors and predecessors, 
which is in agreement with the notion of state transitions 
in Petri nets. When there are no resource conflicts
between two partially ordered threads, they can take
place in parallel (but not necessarily simultaneously), 
and their common predecessors
and successors provide a synchronisation mechanism. 
This is because the common predecessors deliver the 
resources for the parallel threads, and the 
completion of the execution of the threads delivers 
the  resources required by their common successors. 
This view also provides a model of causal flow in the 
execution.

A certain model of concurrency, which is  closely 
related to Petri nets \cite{NPW81}, namely 
\emph{event structures}, captures these ideas: in an event
structure, each instance of a state transition is an event.
The dependencies of events of a concurrent 
system are given with a partial order, and 
non-determinism is captured by a conflict 
relation, which is a symmetric irreflexive relation 
of events. When two events are in conflict, the execution 
of one of them instead of the other determines a different 
state space ahead.

\begin{definition}[labelled event structures]
A labelled event structure (LES) is a structure 
$(E, \leq, \#, \mathcal{L}, \ell)$, where 
\begin{itemize}
%%%%%%%%%%%%%%%%%%%%%%%%%%%%%%%%%%%%%%%%%%%%%%%%%%%%%%%%%%%
%%%%%%%%%%%%%%%%%%%%%%%%%%%%%%%%%%%%%%%%%%%%%%%%%%%%%%%%%%%
\item[(i.)]
$E$ is a set of events;
%%%%%%%%%%%%%%%%%%%%%%%%%%%%%%%%%%%%%%%%%%%%%%%%%%%%%%%%%%%
%%%%%%%%%%%%%%%%%%%%%%%%%%%%%%%%%%%%%%%%%%%%%%%%%%%%%%%%%%%
\item[(ii.)]
$\leq \subseteq E^2$ is a partial order such that for 
every $\mathsf{e} \in E$ the set 
$\{ \mathsf{e}' \in E \, | \, \mathsf{e}' \leq \mathsf{e} \}$ is finite;
%%%%%%%%%%%%%%%%%%%%%%%%%%%%%%%%%%%%%%%%%%%%%%%%%%%%%%%%%%%
%%%%%%%%%%%%%%%%%%%%%%%%%%%%%%%%%%%%%%%%%%%%%%%%%%%%%%%%%%%
\item[(iii.)]
the conflict relation $\# \, \subseteq E^2$ is a symmetric and 
irreflexive relation such that if $\mathsf{e} \, \# \, \mathsf{e}'$ and 
 $\mathsf{e'} \, \leq \, \mathsf{e}''$ 
then $\mathsf{e} \, \# \, \mathsf{e}''$ for every
$\mathsf{e}, \mathsf{e'}, \mathsf{e}'' \in E$;
%%%%%%%%%%%%%%%%%%%%%%%%%%%%%%%%%%%%%%%%%%%%%%%%%%%%%%%%%%%
%%%%%%%%%%%%%%%%%%%%%%%%%%%%%%%%%%%%%%%%%%%%%%%%%%%%%%%%%%%
\item[(iv.)]
$\mathcal{L}$ is a set of labels;
%%%%%%%%%%%%%%%%%%%%%%%%%%%%%%%%%%%%%%%%%%%%%%%%%%%%%%%%%%%
%%%%%%%%%%%%%%%%%%%%%%%%%%%%%%%%%%%%%%%%%%%%%%%%%%%%%%%%%%%
\item[(v.)]
$\ell: E \rightarrow \mathcal{L}$ is a labelling function.
%%%%%%%%%%%%%%%%%%%%%%%%%%%%%%%%%%%%%%%%%%%%%%%%%%%%%%%%%%%
%%%%%%%%%%%%%%%%%%%%%%%%%%%%%%%%%%%%%%%%%%%%%%%%%%%%%%%%%%%
\end{itemize}
\end{definition}

There are standard techniques in the literature for obtaining a 
labelled event structure from transition systems and 
from other models of concurrency \cite{WN95}. 
In particular, in \cite{Kah09} we present 
an application of these techniques for obtaining a labelled event 
structure from the multiplicative exponential linear logic 
encodings of Petri nets.
We can analogously carry these ideas to the setting of the 
process models: for every model $\mathcal{M}$,
we use the operational semantics of the stochastic 
$\pi$ calculus, given with the reduction rules in 
Definition \ref{spi:reduction}, as a transition system.
This way, we apply the techniques presented in \cite{Kah09} to 
obtain the $\LES{\mathcal{M}}$ for the model $\mathcal{M}$.

\begin{example}
\label{figure:les}
We obtain the event structure depicted in Figure \ref{figure:Petri}
for the model in Example \ref{example:diamond}.  
\end{example}

\begin{definition}
Given a LES $(E, \leq, \#, \mathcal{L}, \ell)$, for any event $\mathsf{e} \in E$,
$\causes{\mathsf{e}}$ denotes the set $\{ \mathsf{e}' \in  E | \mathsf{e}' < \mathsf{e} \}$
of \emph{causes} of event $\mathsf{e}$.
We say that $\mathcal{C} \subseteq E$ 
is a \emph{configuration} if and only if 
($i.$) for all 
$\mathsf{e} \in \mathcal{C}$ we have 
that $\causes{\mathsf{e}} \subset \mathcal{C}$;
($ii.$) for all $\mathsf{e}, \mathsf{e}' \in \mathcal{C}$, 
it is not the case that $\mathsf{e} \, \# \, \mathsf{e}'$.
\end{definition}

\begin{example}
Consider the event structure given in Figure \ref{figure:Petri}. The sets 
$\{ \mathsf{e}_1, \mathsf{e}_2, \mathsf{e}_3, \mathsf{e}_4\}$ and 
$\{ \mathsf{e}_1, \mathsf{e}_2,\mathsf{e}'_2\}$ are configurations.
\end{example}

In \cite{Kah09}, we have given an algorithm for extracting  
configurations from transition histories of Petri nets. 
Below, we analogously carry 
these ideas to the setting of the process models being 
discussed here: the CTMCs in \textsf{SPiM} treat the processes as  
resources in the sense that processes representing species are 
consumed and new ones are produced as a result of a reduction with \textsf{SPiM}.
`Resource consciousness' is the crucial 
ingredient that delivers the LES 
semantics in \cite{Kah09}. We can thus extract a 
partial order that delivers the causal dependencies 
of the transitions  of a particular simulation 
as sketched in Figure \ref{figure:diamond}. 
Let us first collect some 
definitions that we need to formally describe 
this procedure on \textsf{SPiM} simulations. 

\begin{definition}[simulation state]
Let $\mathcal{M} = (\mathcal{I},\mathcal{T})$ be a model 
with the reactions $\mathsf{r}_1, \ldots, \mathsf{r}_n$.
A \emph{simulation state} of $\mathcal{M}$, denoted with $\mathcal{Z}$, 
is a multiset of species such that 
each species in $\mathcal{Z}$ is parameterised with  
a label $\mathsf{k} \in \mathbb{N}^+$ denoting an id, 
a label $p,q \in \{ \mathsf{init}, \mathsf{r}_1, \ldots, \mathsf{r}_n\}$ 
denoting the reaction that created this species,  
and $t \in \mathbb{R}^+$ denoting its creation time.
The \emph{initial simulation state} $\mathcal{Z}_0$
for the initial state $\mathcal{I}$ of a model is 
obtained from $\mathcal{I}$ by assigning a \emph{unique} label 
$\mathsf{k} \in \mathbb{N}^+$ to each species and marking each species 
with $\mathsf{init}$ and the time $0$.
\end{definition}

In \textsf{SPiM}, we can implement the notion of a simulation 
state by decorating each process with an integer valued \verb+id+, which 
is communicated during the reduction of the process to its continuation. 

\begin{example}
\label{example:initial:simulation:state}
Consider the \textsf{SPiM} program given in Figure \ref{figure:SPiM:model:id},
which is obtained from the program in Figure \ref{figure:spim:model} by assigning an
\verb+id+ to each process. Here, we have initially four instances of the process $A$.
Thus, $\mathcal{Z}_0 = 
\multiset{ A(1, \mathsf{init}, 0), \, A(2, \mathsf{init}, 0), \, 
             A(3, \mathsf{init}, 0), \, A(4, \mathsf{init}, 0)  }$ 
is the initial simulation state for the 
initial state of the model implemented in Figure \ref{figure:SPiM:model:id}.
\end{example}

\begin{figure}[t]
\begin{multicols}{3}
{\footnotesize
\begin{verbatim}
directive sample 1000.0
directive plot A(); P(); 
               B(); C(); D() 

val r1 = 1.0    val r2 = 1.0      
val r3 = 1.0    val r4 = 1.0      

new ch@r4:chan()
let A(id:int) = delay@r1; 
                  (P(id) | P(id))

and P(id:int) = do delay@r2; B(id)
                or delay@r3; C(id)

and B(id:int) = ?ch; D(id)

      and C(id:int) = !ch; ()

      and D(id:int) = ()

      run ( A(1) | A(2) | 
            A(3) | A(4) )
\end{verbatim}
}
\end{multicols}
\caption{A \textsf{SPiM} implementation of the model given in 
Example \ref{example:diamond} where each process is assigned an 
id as a parameter that is communicated during reductions.}
\label{figure:SPiM:model:id}
\end{figure}

\begin{definition}[simulation trajectory]
Let $\mathcal{M} = (\mathcal{I},\mathcal{T})$ be a model. 
A \emph{simulation trajectory} 
$\mathcal{S}$ on ${\mathcal{M}}$ is a list of 
structures of the form $(\mathsf{t}, \mathsf{L}, \mathsf{R})$, 
called  \emph{transition instances}, 
where $\mathsf{t} \in \mathbb{R}^+$ is the time stamp of the transition. 
$\mathsf{L}$ and $\mathsf{R}$ are multisets of 
species that we call \emph{consumed} and \emph{produced} species, respectively.
We assume that for each pair $(\mathsf{L},\mathsf{R})$ 
there is a reaction $\mathsf{r}$ in $\mathcal{T}$ and
the species on the left-hand-side and  right-hand-side of $\mathsf{r}$, 
respectively, are exactly those in $\mathsf{L}$ and $\mathsf{R}$. 
In this case, we say that $(\mathsf{t},\mathsf{L},\mathsf{R})$ 
\emph{is an instance of} $\mathsf{r}$. 
We use $[]$ to denote the empty trajectory and
the operator `$\bullet$' to denote the concatination of a 
transition with a simulation trajectory, e.g., 
$\mathcal{S} = (\mathsf{t},\mathsf{L},\mathsf{R}) \bullet \mathcal{S}'$.   
\end{definition}

We use the simulation trajectories for representing  
the outcome of simulations with  \textsf{SPiM}
models.

\begin{example}
\label{example:simulation:trajectory}
Consider the simulation output given in Figure \ref{figure:trace:id}. 
We denote this output as the simulation trajectory $\mathcal{S}$, 
where $\mathcal{S}'$ is a simulation trajectory that contains the 
transitions after the time point $1.0$. 
$$
\begin{array}{rl}
\mathcal{S} =  \; &
(0.362
%742117504
,   \multiset{A(4)}, \multiset{P(4), P(4)}) \; \bullet 
(0.403
%041278461
,   \multiset{P(4)}, \multiset{B(4)}) \; \bullet 
(0.533
%386757223
,   \multiset{P(4)}, \multiset{C(4)}) \; \bullet
\; \, \mathcal{S}'
%,  
%\\
%&
%(1.38225471287,    \multiset{A(1)}, \multiset{P(1), P(1)}), &
%(1.45347041154,    \multiset{A(2)}, \multiset{P(2), P(2)}), 
%\\
%&
%(1.51272758278,    \multiset{A(3)}, \multiset{P(3), P(3)}), &
%(1.52969767363,    \multiset{P(3)}, \multiset{B(3)}) \; ] 
%, 
%\\
%&
%(1.53959162309,    \multiset{P(2)}, \multiset{B(2)}), &
%(1.61357605349,    \multiset{P(3)}, \multiset{B(3)}) 
%\\
%&
%(1.62828014739,    \multiset{P(1)}, \multiset{B(1)}) &
%(1.74376163407,    \multiset{P(2)}, \multiset{C(2)}) \;\; 
%\\
%&
%(1.93175167917,    \multiset{P(1)}, \multiset{C(1)}) &
%(2.09975124488,    \multiset{B(4), C(1)}, \multiset{D(4)}) &
%(2.10580589848,    \multiset{B(3), C(4)}, \multiset{D(3)}) &
%(2.14571167381,    \multiset{B(2), C(2)}, \multiset{D(2)}) \;
%\;]
\end{array}
$$
\end{example}
\begin{figure}
\begin{multicols}{2}
{\footnotesize
\begin{verbatim}
     0.  --> A(4) A(3) A(2) A(1) 
     0.362742117504 A(4)  --> P(4) P(4) 
     0.403041278461 P(4)  --> B(4) 
     0.533386757223 P(4)  --> C(4) 
     1.38225471287 A(1)  --> P(1) P(1) 
     1.45347041154 A(2)  --> P(2) P(2) 
     1.51272758278 A(3)  --> P(3) P(3) 
     1.52969767363 P(3)  --> B(3) 
     1.53959162309 P(2)  --> B(2) 
     1.61357605349 P(3)  --> B(3) 
     1.62828014739 P(1)  --> B(1) 
     1.74376163407 P(2)  --> C(2) 
     1.93175167917 P(1)  --> C(1) 
     2.09975124488 B(4) C(1)  --> D(4) 
     2.10580589848 B(3) C(4)  --> D(3) 
     2.14571167381 B(2) C(2)  --> D(2) 
\end{verbatim}
}
\end{multicols}
\caption{A \textsf{SPiM} output of a simulation with the model given 
with the code in Figure \ref{figure:SPiM:model:id}.}
\label{figure:trace:id}
\end{figure}

\begin{definition}[simulation configuration]
Let $\mathcal{M} = (\mathcal{I},\mathcal{T})$ be a model 
with the reactions $\mathsf{r}_1, \ldots, \mathsf{r}_k$. 
The \emph{simulation configuration} for a simulation trajectory 
$\mathcal{S}$ and an initial simulation state $\mathcal{Z}_0$,
denoted by $\mathsf{C}_{\mathcal{S},\mathcal{Z}_0}\,$, 
is the structure $\mu(\mathcal{S},\mathcal{Z}_0)$,
where the recursive 
function $\mu$ on the pairs of simulation trajectories 
and simulation states is defined as follows. 
\begin{itemize}
%%%%%%%%%%%%%%%%%%%%%%%%%%%%%%%%%%%%%%%%%%%%%%%%
%%%%%%%%%%%%%%%%%%%%%%%%%%%%%%%%%%%%%%%%%%%%%%%%
\item
if $\mathcal{S} = (\mathsf{t},\mathsf{L},\mathsf{R}) \bullet \mathcal{S}'$
such that 
$\mathsf{L} = \multiset{A(m)}$, 
$\mathsf{R} = \multiset{P_1(m),\ldots, P_k(m)}$,  
%%%%%%%%%%%%%%%%%%%%%%%%%%%%%%%%%%%%%%%%%%%%%%%%
\begin{itemize}
\item
there is a reaction $\mathsf{r} \in \{ \mathsf{r}_1, \ldots, \mathsf{r}_k \}$ such that 
$(\mathsf{t},\mathsf{L},\mathsf{R})$ is an instance of $\mathsf{r}$, and
\item
for transition label $p$ and time label $t'$, $A(m,p,t') \in \mathcal{Z}$, 
\end{itemize}
$$
\begin{array}{l}
\textrm{then } 
\mu(\mathcal{S},\mathcal{Z}) =  
\{ \langle(m,\ p,\, t'), (m,\, \mathsf{r},\, \mathsf{t}) \rangle \} \;\\[2pt]
\qquad \qquad \qquad \qquad \cup \; \,
\mu(\mathcal{S}',(\mathcal{Z} \musminus \multiset{A(m,p,t')}) \; \muscup \; 
\multiset{P_1(m,\mathsf{r},\mathsf{t}),\ldots, P_k(m,\mathsf{r},\mathsf{t})}).
\end{array}
$$
%%%%%%%%%%%%%%%%%%%%%%%%%%%%%%%%%%%%%%%%%%%%%%%%
%%%%%%%%%%%%%%%%%%%%%%%%%%%%%%%%%%%%%%%%%%%%%%%%
\item
if $\mathcal{S} = (\mathsf{t},\mathsf{L},\mathsf{R}) \bullet \mathcal{S}'$
such that 
$\mathsf{L} = \multiset{A(m_1),B(m_2)}$, 
$\mathsf{R} = \multiset{P_1(x),\ldots, P_k(x)}$, $x = m_1 \lor x = m_2$,  
%%%%%%%%%%%%%%%%%%%%%%%%%%%%%%%%%%%%%%%%%%%%%%%%
\begin{itemize}
\item
there is a reaction $\mathsf{r} \in \{ \mathsf{r}_1, \ldots, \mathsf{r}_k \}$ such that 
$(\mathsf{t},\mathsf{L},\mathsf{R})$ is an instance of $\mathsf{r}$, and 
\item
for transition labels $p,q$ and time labels $t', t''$, 
$A(m_1,p,t') \in \mathcal{Z}$ and $A(m_2,q,t'') \in \mathcal{Z}$, 
\end{itemize}
$$
\begin{array}{l}
\textrm{then } 
\mu(\mathcal{S},\mathcal{Z}) =  
\{ \langle (m_1,\, p, \,  t'), (x,\, \mathsf{r},\, \mathsf{t})\rangle, \, 
   \langle (m_2, \, q,\, t''), (x,\, \mathsf{r},\, \mathsf{t})\rangle \} \; \\[2pt]
\qquad \qquad \qquad \qquad \cup \; \,
\mu(\mathcal{S}',(\mathcal{Z} \musminus 
\multiset{A(m_1,p,t'), \, A(m_2,q,t'') }) \; \muscup \; 
\multiset{{P_1}(x,\mathsf{r},\mathsf{t}),\ldots, {P_k}(x,\mathsf{r},\mathsf{t})}).
\end{array}
$$
%%%%%%%%%%%%%%%%%%%%%%%%%%%%%%%%%%%%%%%%%%%%%%%%
%%%%%%%%%%%%%%%%%%%%%%%%%%%%%%%%%%%%%%%%%%%%%%%%
\item
Otherwise $\mu(\mathcal{S},\mathcal{Z}) = \emptyset$.
\end{itemize}
\end{definition}

\begin{figure}[t]
$$
\includegraphics[width=120mm]{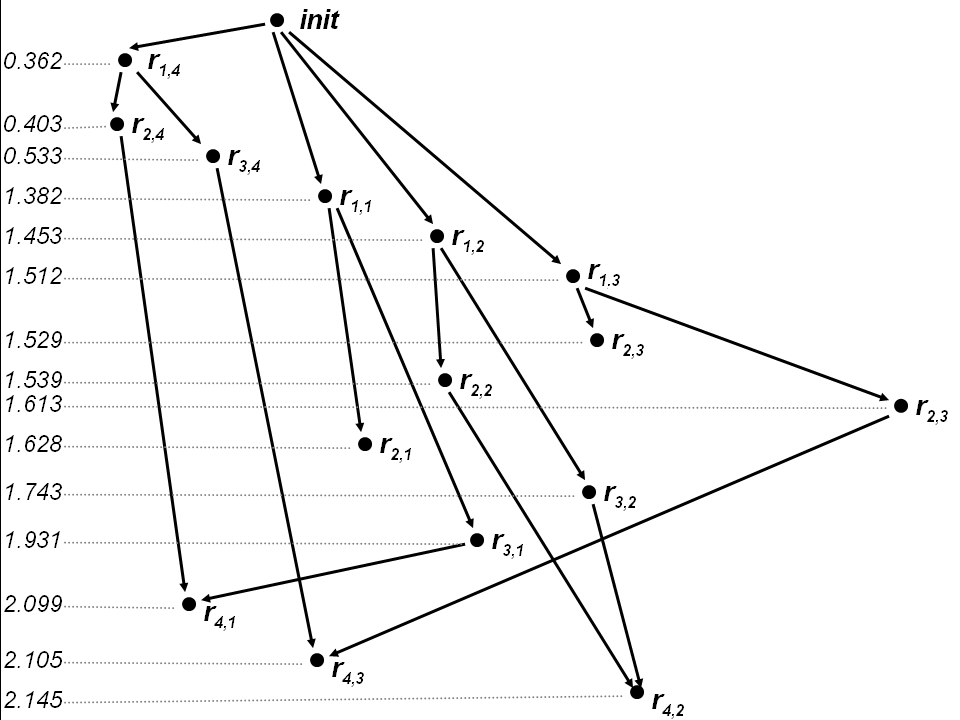}
\!\!\!
\includegraphics[width=40mm]{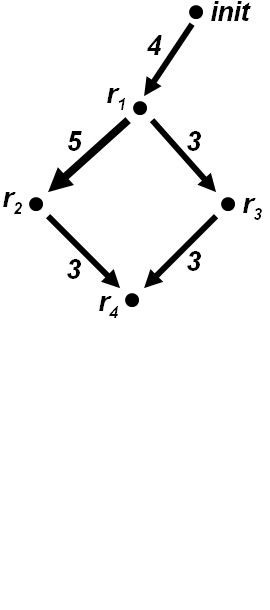}
$$
\caption{The simulation configuration 
obtained from the simulation trajectory delivered 
by the simulation output in Figure \ref{figure:trace:id}, and the 
graphical representation of the flux configuration 
obtained from it by merging the events that are instances of 
the same reactions. 
On the left, 
each node $(x,\mathsf{r}_i,t)$ is abbreviated with $\mathsf{r}_{i,x}$.
On the right, 
the thickness of the arrows are proportional with the 
weights of the edges. }
\label{figure:simulation:configuration}
\end{figure}

\begin{example}
Consider the model $\mathcal{M}$ with the \textsf{SPiM} code in 
Figure \ref{figure:SPiM:model:id}, the initial simulation state 
$\mathcal{Z}_0$ given in Example \ref{example:initial:simulation:state}
and the simulation trajectory $\mathcal{S}$ given in 
Example \ref{example:simulation:trajectory}. 
We get the simulation configuration $\mathsf{C}_{\mathcal{S},\mathcal{Z}_0}$ 
depicted in Figure \ref{figure:simulation:configuration}.
\end{example}

The following proposition is a special case of a more general discussion presented in \cite{Kah09}. 

\begin{proposition}
Let $\mathcal{M}$ be a model with an initial simulation state $\mathcal{Z}_0$
and a simulation trajectory $\mathcal{S}$. The transitive reflexive closure of 
the simulation configuration $\mathsf{C}_{\mathcal{S},\mathcal{Z}_0}$ 
is a configuration in $\LES{\mathcal{M}}$.
\end{proposition}

\section{Quantitative Flux Analysis on Models}
\label{section:quantitative}

A simulation configuration contains the information on the causal 
dependencies of the transitions of a simulation. 
These causal dependencies 
%of the transitions 
are expressed as a partial order 
of individual events. However, despite the partial order representation of 
the causal flow, it remains possible to read the 
total order of the transitions given in the simulation
%in a simulation configuration 
with respect to the time stamps of these events.

In this respect, a simulation configuration is a representation of a simulation, 
where the previously implicit information is made explicit.  
Although such a representation is rich in 
information, it is difficult to make observations 
about the quantitative behaviour of the modelled system. Thus, a simulation 
configuration can be seen as raw data that needs to be processed further for analysis.

The idea here is to apply various transformations 
to the partially ordered structures of simulation configurations 
to extract useful information on the 
dynamics of the system. There are a variety of 
possible choices for such transformations. In the following, we consider 
a transformation, which can provide a means for \emph{quantitative flux 
analysis} in the stochastic setting of process algebra models. In this 
transformation,  we merge the events that are instances of the same reaction 
to the same node, while counting the number of edges that collapse
to a single edge. This way, we assign weights to the edges of the graph 
and obtain a cyclic graph with weights. 

\begin{definition}[flux configuration]
Let $\mathsf{C}_{\mathcal{S},\mathcal{Z}_0}$ be a simulation configuration.
The flux configuration of $\mathsf{C}_{\mathcal{S},\mathcal{Z}_0}$, 
denoted by $\mathsf{F}_{\mathcal{S},\mathcal{Z}_0}$, 
is the set that contains the triples of the form $(p,q,n)$
such that 
$(p,q,n) \in \mathsf{F}_{\mathcal{S},\mathcal{Z}_0}$
if and only if 
there are $n$ number of pairs 
$\langle(i,\, p,\, t),(j,\, q,\, t')\rangle 
\in \mathsf{C}_{\mathcal{S},\mathcal{Z}_0}$ 
for all the possible $i, j, t, t'$.
\end{definition}

\begin{example}
Consider the simulation configuration 
$\mathsf{C}_{\mathcal{S},\mathcal{Z}_0}$ 
depicted on the left-hand-side 
in Figure \ref{figure:simulation:configuration}.  
We obtain the flux configuration
$
\mathsf{F}_{\mathcal{S},\mathcal{Z}_0} = 
\{ 
( \mathsf{init}, \mathsf{r}_1, 4),  
( \mathsf{r}_1, \mathsf{r}_2, 5), 
( \mathsf{r}_1, \mathsf{r}_3, 3), 
( \mathsf{r}_2, \mathsf{r}_4, 3),
( \mathsf{r}_3, \mathsf{r}_4, 3)
\}
$
which is depicted as the graph on 
the right-hand-side in Figure  
\ref{figure:simulation:configuration},
where the numbers on the edges are the 
weights of the edges. 
\end{example}

In comparison to the simulation trajectories and the simulation configurations,
the flux configurations provide a more intelligible representation of the causal 
flow in a simulation from a quantitative point of view. In particular, in a flux 
configuration, the quantitative information on the  flux of the species between 
reactions in the system is highlighted with respect to the simulation. This way 
it becomes possible to make observations on the system that are impossible, 
for example, by counting the number of species or reactions.
However, it is important to note that there can be various choices to be made 
while processing a simulation configuration.  For example, when we consider the 
transformation proposed above, we could have considered only causally independent 
events while merging them. 
While this criteria would not alter the graph in Figure \ref{figure:simulation:configuration},
it would deliver different graphs for the model that we discuss in the next section.

\section{A Biological Model: Rho GTP-binding Proteins}

In this section, we apply the ideas discussed above to a 
model of a biological system, that is, 
\emph{Rho GTP-binding proteins}  \cite{JH05,GP06,CCGKP08}.  

\begin{figure}[t]
$$
\begin{array}{c}
\includegraphics[width=80mm]{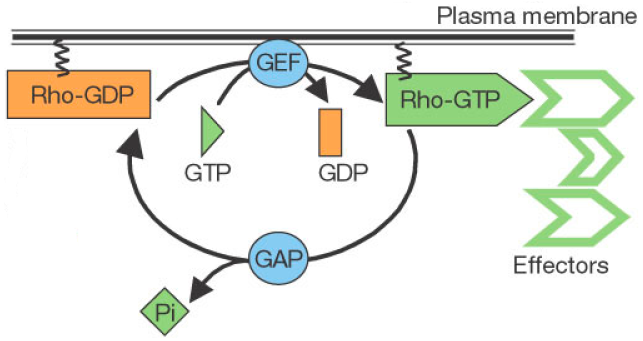}
\end{array}
\qquad \quad
\begin{array}{c}
\includegraphics[width=60mm]{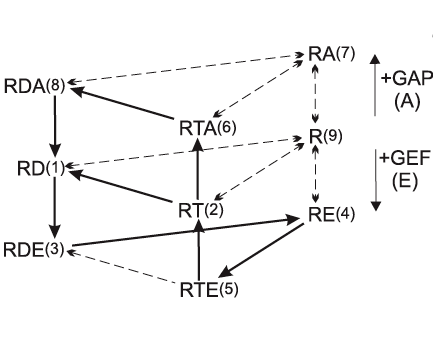}
\end{array}
$$
\caption{Left: Rho GTP-binding protein cycle, 
adapted with permission from Macmillan Publishers 
Ltd: \emph{Nature} \cite{EH02}, copyright 2002.
Right: the structure of the Rho 
GTP-binding proteins model given in \cite{GP06}.
In the graph, \textsf{R} denotes the Rho GTP-binding
protein, whereas \textsf{RD} and \textsf{RT} denote its GDP and GTP 
bound forms respectively. \textsf{A} and \textsf{E} denote GAP and GEF, 
respectively. Thus, \textsf{RDE}, for example, denotes the protein complex 
formed by \textsf{RD} and \textsf{E}.
}
\label{figure:rho:gtp:model}
\end{figure}

The Rho GTP-binding proteins constitute a distinct family within the 
super-family of Ras-related small GTPases with twenty-two identified 
mammalian members, including Rho, Rac and Cdc42 \cite{EH02}. 
Rho GTP-binding proteins serve as molecular 
switches in various subcellular activities, regulating a variety of 
cell functions, including actin organisation and cell shape, 
cell adhesion, cell motility, membrane trafficking and gene 
expression \cite{BSB07}. Their role can be perceived as regulating 
the transmission of an incoming signal further to effectors
in a molecular module by cycling between inactive and active states, 
depending on being GDP or GTP bound, respectively. As depicted in 
Figure \ref{figure:rho:gtp:model}, GDP/GTP cycling is regulated by guanine 
nucleotide exchange factors (GEFs) that promote the GDP dissociation 
and GTP binding, whereas GTPase-activating proteins (GAPs) have the 
opposite effect and stimulate the hydrolysis of Rho-GTP into Rho-GDP.
In the active GTP-bound state, Rho proteins interact with and activate 
downstream effectors.

In \cite{GP06}, Goryachev and Pokhilko give an ordinary differential 
equations (ODE) model of the Rho GTP-binding proteins. The structure of 
this ODE model is depicted on the right-hand-side in 
Figure \ref{figure:rho:gtp:model}:
the three forms of the Rho protein 
(GDP-bound \textsf{RD}, 
 GTP-bound \textsf{RT},
 and nucleotide free \textsf{R}) in the middle layer 
form complexes with GEF (\textsf{E})
in the bottom layer
and with GAP (\textsf{A}) in the top layer.
All the reactions except GTP hydrolysis 
($2 \rightarrow 1$, 
$6 \rightarrow 8$,
$5 \rightarrow 3$) are reversible.
In their model, 
Goryachev and Pokhilko use the quantitative
biochemical data on the Cdc42p member of the Rho family proteins. 
This results in an explanation of the experimentally observed rapid cycling 
of the Rho GTP-binding proteins between their GDP-bound off states and 
GTP-bound on states while displaying high activity, measured by the relative 
concentration of the GTP-bound Rho 
proteins ($\mathsf{RT}$ in Figure \ref{figure:rho:gtp:model}). 
The authors argue that GEF and GAP play distinct and 
separable roles in cycling control: the activity of Rho proteins 
is mainly delivered by the activity of the GEF and the turnover 
rate is a function of the GAP concentration. Therefore, to achieve 
a high activity and turnover rate simultaneously,  the concentrations 
of GEF and GAP should be tightly controlled. Moreover, at large 
$\mathsf{E}_0$ and $\mathsf{A}_0$, only a subset of the 
fluxes~\footnote{In \cite{GP06}, the flux $J_{lm}$
for a reaction that 
connects species $l$ and $m$ is defined with respect to the 
species concentrations and the rate constants, e.g., flux $J_{13}$
connecting \textsf{RD} and \textsf{RDE} is 
$J_{13} = k_{13}. \mathsf{RD} . \mathsf{E} - k_{31} . \mathsf{RDE}$
where  $k_{13}$ and $k_{31}$ are the corresponding reaction rates.}
of the original model is significant, while the remaining fluxes 
have negligible values. To test this hypothesis, Goryachev and 
Pokhilko introduce a reduced model and provide a comparison with the original 
model with  respect to the flux vectors that substantiates the claim. 
Goryachev and Pokhilko 
argue that in the efficient regime the operation of the GEF-GAP 
control module is based on two linear pathways: the GEF arm 
$1 \rightarrow 3 \rightarrow 4 \rightarrow 5 \rightarrow 2 \rightarrow 1$
and the GAP arm $2 \rightarrow 6 \rightarrow 8 \rightarrow 1$, which are  
indicated by solid arrows in Figure \ref{figure:rho:gtp:model}.

In \cite{CCGKP08}, Cardelli et al   
give a stochastic $\pi$ calculus model of the Rho GTP-binding 
proteins, which is based on the 
ordinary differential equations model of \cite{GP06}.   
The model in \cite{CCGKP08}, displays an excellent agreement with 
the ODE model of \cite{GP06} with respect 
to the $\mathsf{RT}$ activity on simulations with varying model 
structures and different regimes of initial concentrations.
In the following, we consider this model for flux analysis by 
employing our software tool that implements the ideas above.

\subsection{Rho without Regulators}
\begin{figure}[t]
$$
\begin{array}{c}
\includegraphics[width=65mm]{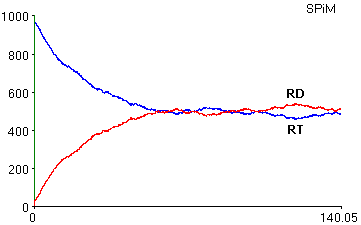}
\end{array}
\qquad 
\begin{array}{c}
\includegraphics[width=80mm]{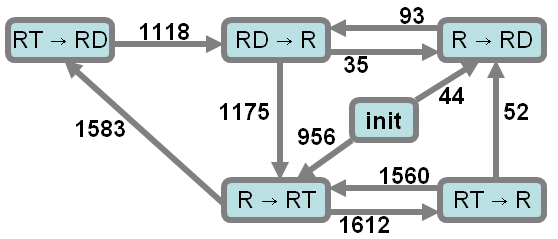}
\end{array}
$$
\caption{A simulation with the reduced model, given with 
the mid layer of the model depicted in 
Figure \ref{figure:rho:gtp:model}, and the 
graph representation of the flux configuration of a 
simulation for the interval $0 < t < 140$.
}
\label{figure:rho:gtp:model:sim}
\end{figure}

At a first step for flux analysis with respect to simulations on 
a stochastic $\pi$ calculus model of the Rho GTP-binding proteins, 
we consider the mid-layer of the model depicted on the right-hand-side 
of Figure \ref{figure:rho:gtp:model}. The \textsf{SPiM} code of the complete 
model is given in Appendix A. For the simulations on the reduced model, 
we run simulations on the complete model where the initial number of 
species for \textsf{A} and \textsf{E} are set as $0$. The initial 
number of $\mathsf{R}$ is set as $1000$.
Because a variation on $\mathsf{R}$
levels does not cause a significant variation in the simulation behaviour 
as it can be observed from the results in \cite{GP06} and \cite{CCGKP08}, 
here we do not consider variations on the $\mathsf{R}_0$.
The plot that displays 
the evolution of the number of species at a simulation that is run until 
the time  $t = 140$ min. is 
depicted~\footnote{The time unit of the model in \cite{GP06} is given in minutes.} 
on the left-hand-side of 
Figure \ref{figure:rho:gtp:model:sim}.

\begin{figure}[b]
$$
\begin{array}{c}
\includegraphics[width=70mm]{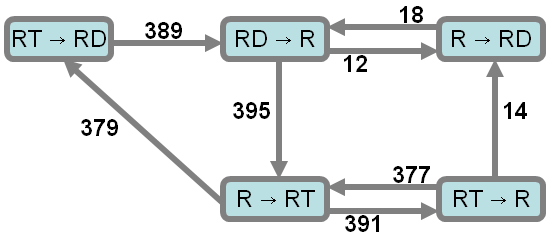}
\end{array}
\quad \leadsto 
\begin{array}{c}
\includegraphics[width=70mm]{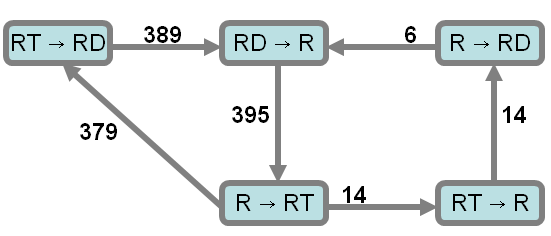}
\end{array}
$$
\caption{
Graph representation of the flux configuration of the 
simulation for the interval $100 < t < 140$.
}
\label{figure:rho:gtp:model:flux:100:140}
\end{figure}

The graph representation of the flux configuration 
for the time interval $0 < t < 140$
of a simulation is given on the right-hand-side 
of the Figure \ref{figure:rho:gtp:model:sim}.
In this plot, we observe that after the start of the simulation, the reactions 
$\mathsf{R} \rightarrow \mathsf{RD}$ and $\mathsf{R} \rightarrow \mathsf{RT}$
are competing for $\mathsf{R}$ and the flux ratio of  
these reactions reflect the bias of the rate constants of these reactions 
towards the reaction $\mathsf{R} \rightarrow \mathsf{RT}$.
%%%%%%%%%%%%%%%%%%%%%%%%%%%%%%%%%%%%%%%%%%%%%%%%%%%%%%%%%%%%%%%%%%%%%%%%%
%%%%%%%%%%%%%%%%%%%%%%%%%%%%%%%%%%%%%%%%%%%%%%%%%%%%%%%%%%%%%%%%%%%%%%%%% 
Because the flux configuration is for the simulation until the time $t =140$, 
the flux of the resources demonstrates the system's accumulation of steady 
state numbers of species from the start of the simulation: the initial 
concentrations are $\mathsf{R}_0 = 1000$,  $\mathsf{RD}_0 = 0$ and $\mathsf{RT}_0 = 0$.
At the end of the simulation we have $\mathsf{R} = 1$, 
$\mathsf{RD} = 503$ and $\mathsf{RT} = 496$. When we examine each 
reaction in the graph with respect to the weights of their incoming 
and outgoing arrows, we observe that the quantities correspond to those 
accumulated at the end of the simulation as demonstrated below and they 
sum up to $1000$.

$$
\footnotesize{
\begin{array}{c}
\left.
\begin{array}{l}
\mathsf{RT} \rightarrow \mathsf{RD}: 1583 - 1118 = 465 \, \mathsf{RD}\\[2pt]
\mathsf{R} \rightarrow \mathsf{RD}: (35 + 44 + 52) - 93 =  38 \, \mathsf{RD}\\[2pt] 
\end{array}
\right\}  503 \mathsf{RD}
\quad
\left.
\begin{array}{l}
\mathsf{RD} \rightarrow \mathsf{R}: (1118 + 93) - (35 + 1175) = 1 \, \mathsf{R} \\[2pt]
\mathsf{RT} \rightarrow \mathsf{R}: 1612 - (1560 + 52) = 0 \, \mathsf{R} \\[2pt]
\end{array}
\right\} 1 \mathsf{R}
\\[16pt]
 \mathsf{R} \rightarrow \mathsf{RT}: (1560 + 956 + 1175) - (1583 + 1612) =  496\, \mathsf{RT}
\end{array}
}
$$

\begin{figure}[t]
$$
\begin{array}{c}
\includegraphics[width=70mm]{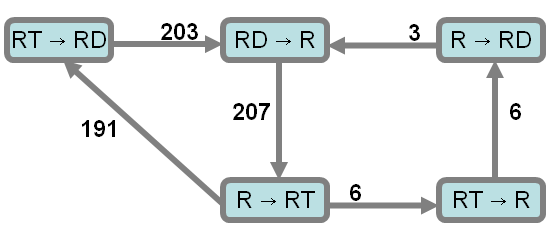}
\end{array}
\quad \quad 
\begin{array}{c}
\includegraphics[width=70mm]{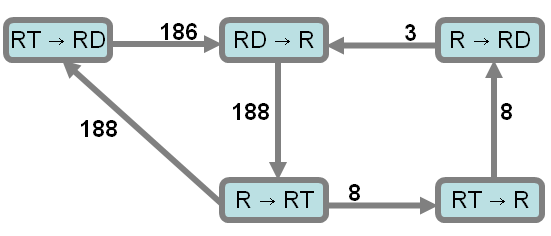}
\end{array}
$$
\caption{
Graph representations of the flux configurations of the 
simulations for the intervals $100 < t < 120$ and $120 < t < 140$.
}
\label{figure:rho:gtp:model:flux:intervals}
\end{figure}

By relying on the data that the cycling period of the system is 
$100$ min \cite{GP06}, we analyse the flux configuration on the same 
simulation for the time interval $100 < t < 140$. This provides 
an examination of the steady-state behaviour of the system.
The graphical representation 
of the flux configuration for this interval 
is given on the left-hand-side of 
Figure \ref{figure:rho:gtp:model:flux:100:140}.
On the right-hand-side of this Figure, we simplify 
this graph by taking the sum of the fluxes for the cases, where 
there are opposite directional fluxes between two reactions. 
For example, we reduce the two arrows between the reactions 
$\mathsf{R} \rightarrow \mathsf{RT}$
and $\mathsf{RT} \rightarrow \mathsf{R}$ to a single arrow.
Below we examine the incoming and outgoing fluxes for each reaction. 

$$
\footnotesize{
\begin{array}{l}
\mathsf{RT} \rightarrow \mathsf{RD}: 379 - 389 = - 10 \, \mathsf{RD}\\[2pt]
\mathsf{R} \rightarrow \mathsf{RD}: 14 - 6 =  8\, \mathsf{RD}
\end{array}
\quad
\begin{array}{l}
\mathsf{RD} \rightarrow \mathsf{R}: 395 - (389 + 6) = 0 \, \mathsf{R}\\[2pt] 
\mathsf{RT} \rightarrow \mathsf{R}: 14 - 14 = 0 \, \mathsf{R} 
\end{array}
\quad
\begin{array}{l}
\mathsf{R} \rightarrow \mathsf{RT}: 395 - (379 + 14) = 2 \,\mathsf{RT}\\[2pt]
\end{array}
}
$$

The observation that the fluxes of the individual reactions do not  
add up to $0$ reflects the fluctuations in the 
stochastic simulation. However, it is important to observe 
the sum $-10 + 8 + 0 + 0 + 2 = 0$ due to the fact that the 
simulation is performed on a closed system. 
In Figure \ref{figure:rho:gtp:model:flux:intervals}, 
we examine the flux configuration of the 
interval  $100 < t < 140$ further by separating it 
into two intervals  $100 < t < 120$ and  $120 < t < 140$. We observe that 
in these smaller intervals, the fluxes of the reactions are closer to $0$.

\subsection{Rho with GEF and GAP}
The behaviour of the model with the regulators  GEF (\textsf{E})
and GAP  (\textsf{A}) in \cite{GP06} is insensitive to the $\mathsf{R}$ 
levels in terms of activity given by the  
$\mathsf{RT}/\mathsf{R}_0$ ratio at the steady states of the simulations. 
The model has an activity maximum with   
the initial concentrations $\mathsf{R} = 1.0 mM$, $\mathsf{E} = 0.776 mM$
and  $\mathsf{A} = 0.66 \mu M$.% 

\begin{figure}[t]
$$
\begin{array}{c}
\includegraphics[width=50mm]{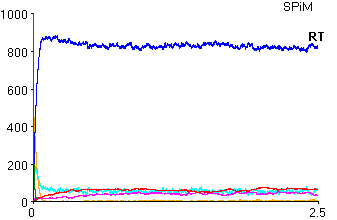}
\quad
\includegraphics[width=50mm]{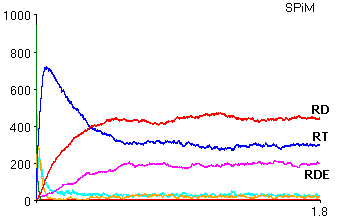}
\quad
\includegraphics[width=50mm]{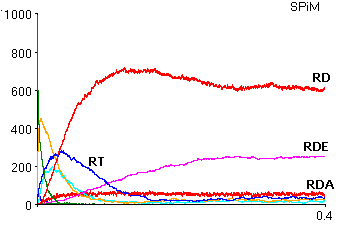}
\end{array}
$$
\caption{Simulation plots with the complete model where
the initial numbers of the species are $\mathsf{R}_0 = 1000$
and $\mathsf{E}_0 = 776$. From left to right, the $\mathsf{A}_0$
value is set as $1$, $10$ and $100$. An increase in  
$\mathsf{A}_0$ in the simulations causes decrease in 
the $\mathsf{RT}$ activity while reducing 
the recovery time.}
\label{figure:rho:gtp:plot:complete}
\end{figure}

%%%%%%%%%%%%%%%%%%%%%%%%%%%%%%%%%%%%%%%%%%%%%%%%%%%%%%%%%
In order to analyse the flux of the system at the high activity regime, 
we run simulations with $\mathsf{R}_0 = 1000$, $\mathsf{E}_0 = 776$
and  $\mathsf{A}_0 = 1$, where we take the closest positive integer number 
value for $\mathsf{A}_0$ so that factoring of the other simulation parameters 
would not be required. This results in a near-maximum activity of approx $0.8$ 
at the steady state with fluctuations due to stochastic simulation. 
A simulation plot with the model in Appendix A with these parameters 
is given on the left-hand-side of Figure \ref{figure:rho:gtp:plot:complete}.
%%%%%%%%%%%%%%%%%%%%%%%%%%%%%%%%%%%%%%%%%%%%%%%%%%%%%%%%%

\begin{figure}[t]
$$
\begin{array}{l}
\mathsf{A}_0 = 1; \, 2.0 < t < 2.5\\[1pt]
\qquad \includegraphics[width=140mm]{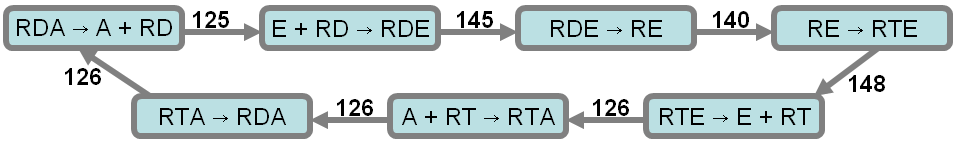}\\[10pt]
%%%%%%%%%%%%%%%%%%%%%%%%%%%%%%%%%%%%%%%%%%%%%%%%%%%%%%%%%%%%%%%%%%%%%%%%%%%%%%%%
\mathsf{A}_0 = 10; \, 1.7 < t < 1.8\\[1pt]
\qquad \includegraphics[width=140mm]{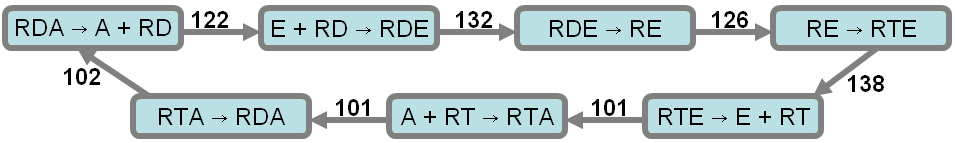}\\[10pt]
%%%%%%%%%%%%%%%%%%%%%%%%%%%%%%%%%%%%%%%%%%%%%%%%%%%%%%%%%%%%%%%%%%%%%%%%%%%%%%%%
\mathsf{A}_0 = 100; \, 0.3 < t < 0.4\\[1pt]
\qquad \includegraphics[width=140mm]{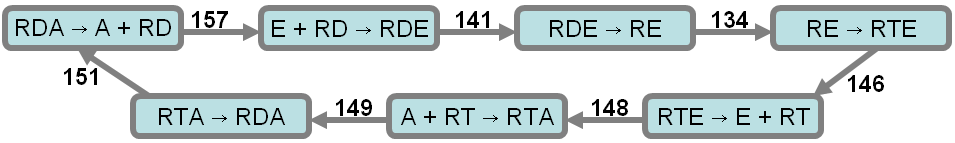}
\end{array}
$$
\caption{The graphs displaying the dominant fluxes with respect to the
flux configurations obtained from  
simulations with $\mathsf{R}_0 = 1000$, $\mathsf{E}_0 = 776$ and 
varying initial $\mathsf{A}_0$ numbers. An increase 
in $\mathsf{A}_0$ increases the turnover rate,  
observed by comparing the ratio of the fluxes and the length 
of the time interval.}
\label{figure:rho:gtp:plot:complete:flux}
\end{figure}

We analyse the steady state behaviour of the system with 
respect to the simulation at this regime. The output 
of our tool, which displays the flux configuration
of the simulation for the time interval $2.0 < t  < 2.5$, 
is given in Appendix B. 
From this flux configuration, we obtain the top-most graph 
in Figure \ref{figure:rho:gtp:plot:complete:flux}, where we omit the 
fluxes which are less than $10 \%$ of the mean of the fluxes. 
This delivers the more dominant flux pathway in the system 
at steady state, given with
$2 \rightarrow 6 \rightarrow 8 \rightarrow 1 \rightarrow 3 \rightarrow 
4 \rightarrow 5 \rightarrow 2$, which is in agreement with
the results of \cite{GP06} with the exception that the 
path  $2\rightarrow 1$ is not included in our graph. 
The presence of this flux in the pathway could be explained 
only with the presence of a flux from the reaction 
$\mathsf{RTE} \rightarrow \mathsf{RT}$ to the
reaction $\mathsf{RT} \rightarrow \mathsf{RD}$. Although
this flux can be read with a weight of $9$ in our flux configuration
(``\verb+22 ==> 9 appears 9 times.+'' in Appendix B), 
it is much smaller in weight in comparison 
to other fluxes that are depicted in the top-most graph in 
Figure \ref{figure:rho:gtp:plot:complete:flux}.
Moreover, we observe that the reaction 
$\mathsf{RTE} \rightarrow \mathsf{RT} + \mathsf{E}$ 
and its reverse reaction
$\mathsf{RT} + \mathsf{E} \rightarrow \mathsf{RTE}$ 
deliver a strong cyclic flux, which cancels itself, 
thus they are not included in the graph 
in Figure \ref{figure:rho:gtp:plot:complete:flux}. 

When we carry this analysis to the simulations where $\mathsf{A}_0$
is increased to $10$ and $100$, we get the simulation plots depicted 
in the middle and right-hand-side of 
Figure \ref{figure:rho:gtp:plot:complete}. 
We observe the same flux pathway patterns for these simulations as 
depicted as the middle and bottom graphs in 
Figure \ref{figure:rho:gtp:plot:complete:flux}. However, we 
observe that an increase in $\mathsf{A}_0$ does not only decrease 
the $\mathsf{RT}$ activity, but also increases the turnover rate. 

It is important to observe that the graphs depicted in 
Figure \ref{figure:rho:gtp:plot:complete:flux} are 
generated from the model, and they reflect the 
predictions of the results in \cite{GP06}. 
Along these lines, Goryachev and Pokhilko argue that 
a subset of the fluxes of the original model is significant
in the actual biological system, 
while the remaining fluxes have negligible values.
Goryachev and Pokhilko substantiate this 
prediction by comparing the reduced 
model with the original model. 
In this respect, the graphs  
given in Figure \ref{figure:rho:gtp:plot:complete:flux}
depict a reduced model and agree with the predictions of \cite{GP06} 
(with the exception of $\mathsf{2} \rightarrow \mathsf{1}$ 
as explained above). However, the graphs
in Figure \ref{figure:rho:gtp:plot:complete:flux}
result from the application of the ideas presented in 
Section \ref{section:causality} and 
Section \ref{section:quantitative}
to the simulations, thus they are generated
automatically, and their delivery does not require 
an analysis of the model or the biological system 
being modelled.

\section{Discussion}
We have shown that the event structures view of simulations discussed above 
provide the means for flux analysis for process algebra models in a 
biologically meaningful manner.  
However, the ideas presented here should provide 
a starting point for further exploration rather than being conclusive. 

The relationship between process models and causality has been studied by various 
authors. In this regard, these previous investigations should provide touchstones 
for further investigations. 
Some of the works along these lines, which also provide further reference pointers, 
are as follows.
In \cite{DFFHK07}, Danos et al draw connections between computational 
models of biological systems, event structures and their causality interpretation, 
while exploiting conflict as a mechanism of inhibition in a signalling pathway. 
The work by Fages and Soliman \cite{FS08}
attacks a similar problem by providing a formal interpretation of the relationship 
between reaction graphs and activation/inhibition graphs used by biologists. 
In \cite{CDPB04}, Curti et al apply the ideas presented in \cite{DP01} to the
$\pi$ calculus models of biological systems where the causality information 
on the modelled system is retrieved by labelling the syntax tree of the 
process expressions.

Topics of future work include the application of the ideas above to 
BlenX \cite{PQR09} language as well as an exploration of the 
relationship with the syntactical approach to causality 
with respect to enhanced operational semantics \cite{DP01}.\\[4pt]
\textbf{Acknowledgements }
We thank  Luca Cardelli for discussions and helpful suggestions, and 
Andrew Philips for his assistance with the $\mathsf{SPiM}$ language.
We thank anonymous referees for valuable comments.

\bibliographystyle{plain}
\bibliography{ozan}

\newpage

\section*{Appendix A}
\begin{verbatim}
directive sample 4.0 
directive plot RD(); R(); RT();
               RDE(); RE(); RTE();
               RDA(); RTA(); RA()

val D = 50.0
val T = 500.0

new a1@1.0:chan()
new a2@1.0:chan()
new a3@1.0:chan()

new e1@0.0054:chan()
new e2@0.0075:chan()
new e3@0.43:chan()


let A() = do ?a1; ()
          or ?a2; () 
          or ?a3; ()

let E() = do ?e1; ()
          or ?e2; ()
          or ?e3; ()

let R(id:int) =  do delay@0.033*D; RD(id)             (* 9 -> 1 *)
                 or delay@0.1*T; RT(id)               (* 9 -> 2 *)
                 or !a3; RA(id)                       (* 9 -> 7 *)
                 or !e3; RE(id)                       (* 9 -> 4 *)

and RD(id:int) = do delay@0.02; R(id)                 (* 1 -> 9 *)
                 or !a1; RDA(id)                      (* 1 -> 8 *)
                 or !e1; RDE(id)                      (* 1 -> 3 *)
           
and RT(id:int) = do delay@0.02; R(id)                 (* 2 -> 9 *)
                 or delay@0.02; RD(id)                (* 2 -> 1 *) 
                 or !e2; RTE(id)                      (* 2 -> 5 *)
                 or !a2; RTA(id)                      (* 2 -> 6 *)

and RE(id:int) = do delay@0.033*D; RDE(id)            (* 4 -> 3 *)
                 or delay@0.1*T; RTE(id)              (* 4 -> 5 *)
                 or delay@1.074; ( R(id) | E() )      (* 4 -> 9 *)

and RDE(id:int) = do delay@0.136; ( RD(id) | E() )    (* 3 -> 1 *)      
                  or delay@6.0; RE(id)                (* 3 -> 4 *) 

and RTE(id:int) = do delay@76.8; ( RT(id) | E() )     (* 5 -> 2 *)
                  or delay@0.02; RDE(id)              (* 5 -> 3 *)
                  or delay@0.02; RE(id)               (* 5 -> 4 *)

and RTA(id:int) = do delay@0.0002; RA(id)             (* 6 -> 7 *)
                  or delay@2104.0; RDA(id)            (* 6 -> 8 *)
                  or delay@3.0; ( RT(id) | A() )      (* 6 -> 2 *)

and RA(id:int) = do delay@0.1*D; RDA(id)              (* 7 -> 8 *)
                 or delay@0.0085*T; RTA(id)           (* 7 -> 6 *) 
                 or delay@500.0; ( R(id) | A() )      (* 7 -> 9 *)

and RDA(id:int) = do delay@500.0; ( RD(id) | A() )    (* 8 -> 1 *)
                  or delay@0.02; RA(id)               (* 8 -> 7 *)
           
let ProduceR(id:int) = if id < 1001 
                 then (R(id) | ProduceR(id+1) )
                 else()
                          
run 776 of E()
run 1 of A()
run 1 of ProduceR(1)
\end{verbatim}

\newpage

\section*{Appendix B}

\begin{multicols}{2}

\begin{flushleft}
The flux analysis output for the simulation with 
$\mathsf{R}_0 = 1000$, $\mathsf{E}_0 = 776$ and $\mathsf{A}_0 = 1$,
for the time interval $2.0 < t < 2.5$.  
\end{flushleft}

{\footnotesize
\begin{verbatim}
Reactions:
----------

1: A R  --> RA  fires 0 times.
2: RA  --> A R  fires 0 times.
3: RTA  --> A RT  fires 0 times.
4: R  --> RD  fires 0 times.
5: RD  --> R  fires 0 times.
6: R  --> RT  fires 0 times.
7: RTE  --> RDE  fires 1 times.
8: RE  --> RDE  fires 2 times.
9: RT  --> RD  fires 9 times.
10: RDE  --> E RD  fires 3 times.
11: RE  --> E R  fires 2 times.
12: E R  --> RE  fires 7 times.
13: RTE  --> RE  fires 2 times.
14: A RD  --> RDA  fires 16 times.
15: A RT  --> RTA  fires 126 times.
16: RTA  --> RDA  fires 126 times.
17: E RD  --> RDE  fires 132 times.
18: RDA  --> A RD  fires 142 times.
19: RT  --> R  fires 6 times.
20: RDE  --> RE  fires 149 times.
21: RE  --> RTE  fires 145 times.
22: RTE  --> E RT  fires 2240 times.
23: E RT  --> RTE  fires 2095 times.


Extracted dependencies:
-----------------------

6 ==> 23  appears 2 times.
7 ==> 20  appears 1 times.
8 ==> 20  appears 3 times.
9 ==> 14  appears 1 times.
9 ==> 17  appears 5 times.
10 ==> 17  appears 2 times.
11 ==> 12  appears 2 times.
12 ==> 21  appears 4 times.
13 ==> 21  appears 1 times.
14 ==> 18  appears 16 times.
15 ==> 16  appears 126 times.
16 ==> 18  appears 126 times.
17 ==> 10  appears 3 times.
17 ==> 20  appears 145 times.
18 ==> 14  appears 15 times.
18 ==> 17  appears 125 times.
19 ==> 12  appears 5 times.
20 ==> 8  appears 2 times.
20 ==> 11  appears 2 times.
20 ==> 21  appears 140 times.
21 ==> 13  appears 1 times.
21 ==> 22  appears 148 times.
22 ==> 9  appears 9 times.
22 ==> 15  appears 126 times.
22 ==> 19  appears 6 times.
22 ==> 23  appears 2093 times.
23 ==> 7  appears 1 times.
23 ==> 13  appears 1 times.
23 ==> 22  appears 2092 times.

==================================
==================================

\end{verbatim}
}
%%%%%%%%%%%%%%%%%
%%%%%%%%%%%%%%%%%

\begin{flushleft}
The flux analysis output for the simulation with 
$\mathsf{R}_0 = 1000$, $\mathsf{E}_0 = 776$ and $\mathsf{A}_0 = 10$,
for the time interval $1.7 < t < 1.8$.  
\end{flushleft}

{\footnotesize
\begin{verbatim}
Reactions:
----------

1: RA  --> RDA  fires 0 times.
2: RTE  --> RDE  fires 0 times.
3: RTA  --> A RT  fires 0 times.
4: R  --> RD  fires 0 times.
5: RTE  --> RE  fires 0 times.
6: RT  --> RD  fires 0 times.
7: RDE  --> E RD  fires 0 times.
8: RD  --> R  fires 2 times.
9: A R  --> RA  fires 1 times.
10: RA  --> A R  fires 1 times.
11: E R  --> RE  fires 10 times.
12: RE  --> RDE  fires 3 times.
13: RT  --> R  fires 1 times.
14: R  --> RT  fires 2 times.
15: RE  --> E R  fires 10 times.
16: RTA  --> RDA  fires 101 times.
17: A RT  --> RTA  fires 101 times.
18: A RD  --> RDA  fires 157 times.
19: E RD  --> RDE  fires 123 times.
20: RTE  --> E RT  fires 255 times.
21: RDE  --> RE  fires 137 times.
22: E RT  --> RTE  fires 134 times.
23: RDA  --> A RD  fires 261 times.
24: RE  --> RTE  fires 139 times.


Extracted dependencies: 
-----------------------

7 ==> 18  appears 1 times.
7 ==> 19  appears 1 times.
8 ==> 9  appears 1 times.
8 ==> 14  appears 1 times.
9 ==> 10  appears 1 times.
10 ==> 11  appears 1 times.
11 ==> 24  appears 13 times.
12 ==> 21  appears 5 times.
13 ==> 14  appears 1 times.
15 ==> 11  appears 9 times.
16 ==> 23  appears 102 times.
17 ==> 16  appears 101 times.
18 ==> 23  appears 159 times.
19 ==> 21  appears 132 times.
20 ==> 13  appears 1 times.
20 ==> 17  appears 101 times.
20 ==> 22  appears 134 times.
21 ==> 12  appears 3 times.
21 ==> 15  appears 10 times.
21 ==> 24  appears 126 times.
22 ==> 20  appears 117 times.
23 ==> 8  appears 2 times.
23 ==> 18  appears 156 times.
23 ==> 19  appears 122 times.
24 ==> 20  appears 138 times.

==================================
==================================

\end{verbatim}
}

%%%%%%%%%%%%%%%%%
%%%%%%%%%%%%%%%%%

\begin{flushleft}
The flux analysis output for the simulation with 
$\mathsf{R}_0 = 1000$, $\mathsf{E}_0 = 776$ and $\mathsf{A}_0 = 100$,
for the time interval $0.3 < t < 0.4$.  
\end{flushleft}

{\footnotesize
\begin{verbatim}
Reactions:
----------

1: RA  --> RTA  fires 0 times.
2: RA  --> RDA  fires 0 times.
3: R  --> RD  fires 0 times.
4: A R  --> RA  fires 0 times.
5: RTA  --> A RT  fires 0 times.
6: R  --> RT  fires 0 times.
7: RT  --> RD  fires 0 times.
8: RDA  --> RA  fires 1 times.
9: RE  --> E R  fires 1 times.
10: RA  --> A R  fires 1 times.
11: RD  --> R  fires 2 times.
12: E R  --> RE  fires 4 times.
13: RE  --> RDE  fires 4 times.
14: RDE  --> E RD  fires 6 times.
15: E RT  --> RTE  fires 22 times.
16: E RD  --> RDE  fires 157 times.
17: RDE  --> RE  fires 143 times.
18: A RT  --> RTA  fires 148 times.
19: RTA  --> RDA  fires 149 times.
20: RE  --> RTE  fires 140 times.
21: RTE  --> E RT  fires 168 times.
22: A RD  --> RDA  fires 2606 times.
23: RDA  --> A RD  fires 2756 times.


Extracted dependencies: 
-----------------------

8 ==> 10  appears 1 times.
9 ==> 12  appears 1 times.
10 ==> 12  appears 1 times.
11 ==> 12  appears 2 times.
12 ==> 20  appears 6 times.
13 ==> 17  appears 2 times.
14 ==> 22  appears 6 times.
15 ==> 21  appears 22 times.
16 ==> 14  appears 6 times.
16 ==> 17  appears 141 times.
17 ==> 9  appears 1 times.
17 ==> 13  appears 4 times.
17 ==> 20  appears 134 times.
18 ==> 19  appears 149 times.
19 ==> 23  appears 151 times.
20 ==> 21  appears 146 times.
21 ==> 15  appears 22 times.
21 ==> 18  appears 148 times.
22 ==> 8  appears 1 times.
22 ==> 23  appears 2605 times.
23 ==> 11  appears 2 times.
23 ==> 16  appears 157 times.
23 ==> 22  appears 2600 times.

==================================
==================================

\end{verbatim}
}
\end{multicols}

\end{document}